\documentclass[12pt]{raa}
\usepackage{graphicx,times}
\usepackage{natbib}
\usepackage{epsfig}
\usepackage{amsmath}
\usepackage{txfonts}
\usepackage{lscape}

\providecommand{\bm}[1]{\mbox{\boldmath$#1$\unboldmath}}
\newcommand{\chandra}{{\it Chandra}}

\begin{document}

\title{\chandra\ Studies of the X-ray Gas Properties of Fossil Systems}
\author{Zhenzhen Qin \inst{1}}

\institute{\inst{1}School of Science, Southwest University of Science and Technology, 59 Qinglong Road, Mianyang, Sichuan, 621010, China; {\it qin\_zhenzhen@hotmail.com}}

\abstract{
We study ten galaxy groups and clusters suggested in the literature to be ``fossil system (FS)'' based on \chandra\ observations. According to the $M_{500}-T$ and $L_{\rm X}-T$ relations, the gas properties of FSs are not physically distinct from ordinary galaxy groups or clusters. We also first study the $f_{\rm gas,~2500}-T$ relation and find that the FS exhibits same as ordinary systems. The gas densities of FSs within $0.1r_{200}$, are $\sim 10^{-3}$ cm$^{-3}$, which is the same order as galaxy clusters. The entropies within $0.1r_{200}$ ($S_{0.1r_{200}}$) of FSs are systematically lower than those in ordinary galaxy groups which is consistent with previous report, but we find their $S_{0.1r_{200}}-T$ relation is more similar to galaxy clusters. The derived mass profiles of FSs are consistent with the Navarro, Frenk, \& White model in $(0.1-1)r_{200}$, and the relation between scale radius $r_{\rm s}$ and characteristic mass density $ta_{\rm c}$ indicates the self-similarity of dark matter halos of FSs. The range of $r_{\rm s}$ and $ta_{\rm c}$ of FSs are also close to those of galaxy clusters. Therefore, FSs share more common characteristics with galaxy clusters. The special birth place of the FS makes it as a distinct galaxy system type.
}

\authorrunning{Zhenzhen Qin}
\titlerunning{\chandra\ Studies of the X-ray Gas Properties of Fossil Systems}
\maketitle

\section{INTRODUCTION}
Jone et al. (2003) defined a fossil group in observational terms as a spatially extended X-ray source with an X-ray luminosity from hot gas of $L_{\rm X,bol} \geq 10^{42}$ ergs $\rm s^{-1}$, and a bound system of galaxies with $\bigtriangleup m_{\rm 12} \geq 2.0$ mag, where $\bigtriangleup m_{\rm 12}$ is the absolute total magnitude gap in R band between the brightest and second-brightest galaxies in the system within half of the (projected) virial radius. Thus, it's a galaxy group dominated by one central luminous giant elliptical galaxy with few, or no, $L^*$ galaxies inside the radius for orbital decay by dynamical friction. There also exist some fossil clusters like RX J1416.5+4018 and AWM 4. We nomenclature such galaxy group or cluster as fossil system (FS) throughout this paper. Such high $ta m_{12}$ is extremely rare in ordinary galaxy groups or clusters (Beers et al. 1995), which is why the FS is supposed as a distinct galaxy system type.

However, the origin and evolution process of the FS are still not well understood and have many discrepancies. There are mainly three scenarios of the FS origin proposed in previous works: [1], the FS may be the end result of galaxy merging within a normal group (Ponman et al. 1994; Jones et al. 2000); [2], the birth place of the FS may be isolated and deficient in $L^*$ galaxies (Mulchaey \& Zabludoff 1999); [3], the FS is a transient yet common phase in the evolution of groups or clusters, ending with the infall of fresh galaxies from the surroundings (von Benda-Beckmann et al. 2008). The discrepancies of its evolution process mainly focus on two points: whether the FS is the descendant of the compact group (Yoshioka et al. 2004; Mendes de Oliveira \& Carrasco 2007), and the relation between the FS and the galaxy cluster (Khosroshahi et al. 2006; Mendez-Abreu et al. 2012). Simulations indicate FSs are assembled a large fraction of their mass at high redshifts, which means FSs are formed earlier than ordinary systems (Dariush et al. 2007; Dariush et al. 2010). Much effort has been devoted to understand the origin and evolution of the FSs via their optical properties, such as the photometric luminosity function, the stellar population ages and the isophotal shapes of brightest galaxies in FSs (Adami et al. 2012; Aguerri et al. 2011; Khosroshahi et al. 2006). There are also some systematic X-ray studies on hot gas properties of FSs (Khosroshahi et al. 2007; Miller et al. 2012).

In this paper, we analyze the \chandra\ archive data of ten FSs to determine their X-ray characteristics, including temperatures, masses, luminosities, gas fractions and entropies. The scaling relations between them are used to infer the gas accretion history and heating process in FSs. We also study dark matter halo structures of FSs.

We describe the sample selection criteria and data analysis in ${\S}$ 2 \& ${\S}$ 3, respectively. ${\S}$ 4, ${\S}$ 5 and ${\S}$ 6 present results of our sample study and discussion the physical implications. We summarize our work in ${\S}$ 7. We assume $H_{\rm 0}=70$ km s$^{-1}$ Mpc$^{-1}$, a flat universe for which $\Omega_{\rm m} = 0.3$ and $\Omega_{\Lambda} = 0.7$, and adopt the solar abundance standards of Grevesse and Sauval (1998), where the iron abundance relative to hydrogen is $3.16\times10^{-5}$ in number. Unless stated otherwise, all quoted errors are derived at the 68\% confidence level.

\section{SAMPLE AND DATA PREPARATION}

We construct our sample of ten FSs (including eight fossil groups and two fossil clusters) with $z \leq 0.4$, which chosen from confirmed FSs in previous works. Six of them are observed by the S3 CCD of the \chandra\ advanced CCD imaging spectrometer (ACIS) instrument, and the rest four are observed by the ACIS I CCDs. We list basic properties of the sample in Table \ref{table_sample}, which are arranged in the orders of Object names(col. [1]), right ascension and declination coordinates (J2000) of the FS optical centroids (col.[2] \& col.[3]), redshifts (col.[4]), detectors (col.[5]), exposure times (col.[6]), \chandra\ Observational ID (col.[7]) and reference(col.[8]).

\begin{table}[h!!!]
\caption{FSs in Our Sample}\label{table_sample}
\begin{center}
\tiny
\begin{tabular}{clccccrcc}
\hline\hline
Object name&RA$^b$&DEC$^b$&Redshift&Detector&Exp(ks)&ObsID&Reference$^c$\\
\hline
AWM 4 $^a$ & 16h04m57.0s & +23d55m14s & 0.032 & ACIS-S & 75 & 9423 & 1  \\  
ESO 306017 & 05h40m06.7s & -40d50m11s & 0.036 & ACIS-I & 14 & 3188 &2\\ 
NGC 1132 & 02h52m51.8s & -01d16m29s & 0.023 & ACIS-S & 40 & 3576 &3\\ 
NGC 1550 & 04h19m37.9s & +02d24m36s & 0.012 & ACIS-I & 10 & 3186 & 4 \\ 
NGC 6482 & 17h51m48.8s & +23d04m19s & 0.013 & ACIS-S & 20 & 3218 & 5 \\  
NGC 741 & 01h56m21.0s & +05d37m44s & 0.019 & ACIS-S & 31 & 2223  &  6\\  
RX J1340.6+4018 & 13h40m09.0s & +40d17m43s & 0.171 & ACIS-S & 48 & 3223 & 1  \\
RX J1416.4+2315$^a$ &14h16m26.0s & +23d15m23s & 0.138 & ACIS-S & 14 & 2024 &  1\\
SDSS J0150-1005 & 01h50m21.3s & -10d05m31s & 0.364 & ACIS-I & 27 & 11711 & 7 \\
SDSS J1720+2637 & 17h20m10.0s & +26d37m32s & 0.159 & ACIS-I &26 & 4631 &  7\\ 
\hline\hline
\end{tabular}
\end{center}
\tablecomments{0.9\textwidth}
{
\scriptsize
$^a$ AWM 4 is a poor galaxy cluster, and RX J1416.4+2315 was identified as a fossil cluster (Cypriano et al. 2006). $^b$ Positions of FS optical centroids (J2000). $^c$ References including 1 -- Zibetti et al. (2009), 2 -- Sun et al. (2004), 3 -- Yoshioka et al. (2004), 4 -- Sato et al. (2010), 5 -- Khosroshahi et al. (2004), 6 -- Rasmussen et al. (2007), and 7 -- Santos et al. (2007).
}
\end{table}

All the X-ray data analyzed in this work is acquired with the ACIS S/I CCDs temperature set to be $-120$ $^{\circ}$C. Using the CIAO software (version 3.4), we keep events with ASCA grades 0, 2, 3, 4 and 6, remove all the bad pixels and columns, columns adjacent to bad columns and node boundaries, and then  exclude the gain, CTI, and astrometry corrections. In order to identify possible strong background flare, light curves are extracted from regions sufficiently far away. Time intervals during the count rate exceeding the average quiescent value by 20 percent are excluded.

\section{DATA ANALYSIS}

\subsection{Spectral Analysis}
We utilize the \chandra\ blank-sky template for the ACIS CCDs as the
background. The template is tailored to match the actual pointing. The background spectrum is extracted and processed identically to the source spectrum. Then we rescale the background spectrum by normalizing its high energy end to the corresponding observed spectrum. The corresponding spectral redistribution matrix file (RMFS) and auxiliary response files (ARFS) are created by using the CIAO tool $mkwarf$ and $mkacisrmf$, respectively. All spectra are rebinned to insure at least 20 raw counts per spectral bin to allow $\chi^2$ statistics to be applied. Since the contribution of the hard spectral component is expected to be rather weak, and also to minimize the effects of the instrumental background at higher energies as well as the calibration uncertainties at lower energies, the fittings were restricted in $0.7-7.0$ keV.

\begin{figure}
     \begin{center}
       \includegraphics[width=.8\textwidth]{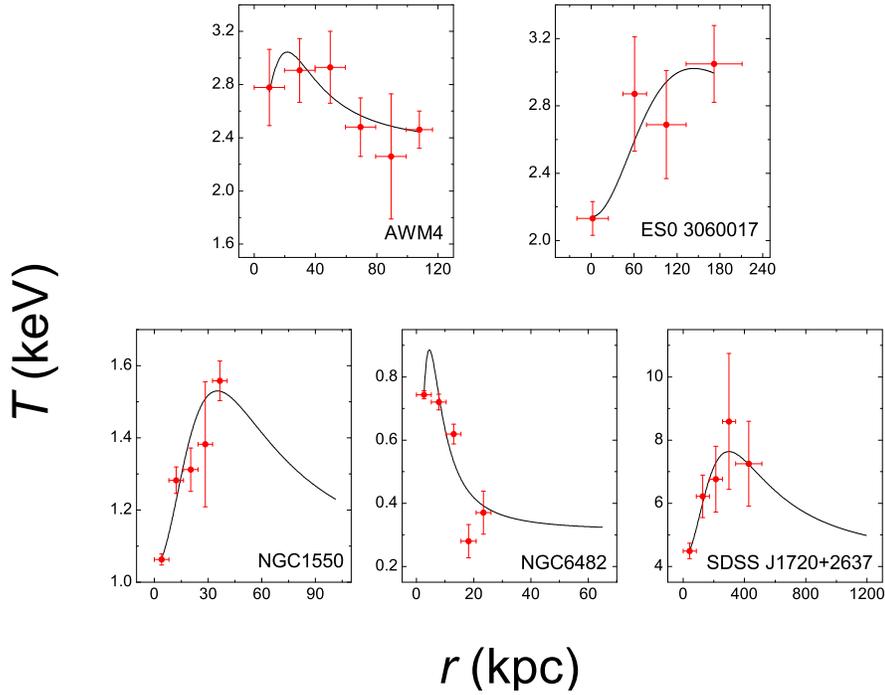}
       \caption{The deprojected temperature profiles of five FSs with sufficient counts, and solid lines are the best-fit deprojected temperature profiles.
       }\label{fig_tprofile}
     \end{center}
   \end{figure}

Due to the sufficient counts numbers, we can extract deprojected temperature profiles of five FSs: AWM 4, ESO 306017, NGC 1550, NGC 6482 and SDSS J1720+2637, where we use the XSPEC model $projct$ to evaluate the influence of the outer spherical shells onto the inner ones. We model the hot gas with absorbed APEC component in every annulus, with an additional power-law component subject to the same absorption to constrain the contribution from unresolved Low Mass X-ray Binaries (LMXBs). The absorption ($N_H$) of each FS is fixed at the Galactic value (Dickey \& Lockman 1990) throughout. The spectra are fitted with XSPEC v.12.3.1x. The best-fit deprojected gas temperature profiles of these five FSs are shown in Fig. \ref{fig_tprofile}. To describe the obtained deprojected best-fit temperature profiles in a smooth form, we adopt the analytic model based on the formulation introduced in Allen et al. (2001),
 
\begin{equation}
T(r) = T_{\rm 0}+T_{\rm 1}{{(r/r_{\rm tc})^\eta}\over{(1+(r/r_{\rm tc})^\eta)^\eta}},
\end{equation}
where $T_{\rm 0}$, $T_{\rm 1}$ and $r_{\rm tc}$ are fitting parameters with fixed $\eta = 2$. {We note that our temperature model differs from that of Allen et al. (2001) in two aspects: (i) it models the radius and temperature in physical units rather than normalized to their value at $R_{2500}$ as the model in Allen et al. (2001); (ii) our model has an extra ``$\eta$" to better describe the temperature drop at outer regions of  galaxy groups and clusters.} The beat-fit $T_{\rm 0}$, $T_{\rm 1}$ and $r_{\rm tc}$ are listed in Table. 2, and the smoothed deprojected temperature profiles are also shown in Fig. \ref{fig_tprofile}. One can see AWM 4 shows nearly isothermal temperature profile, which was reported by (O'Sullivan et al. 2005). NGC 6482 shows a hot core in the temperature profile, {which was ascribed to a possible AGN activity in the group center (Khosroshahi et al. 2004).} The other three FSs show central temperature decreases, which indicate residing cool cores and have no ongoing major merger. 

\begin{table}[h!!!]
\caption{Gas Properties of FSs}\label{table_properties}
\begin{center}
\tiny
\begin{tabular}{clccccrcc}
\hline\hline
Object name & $T_{\rm g}^a$ & $T_{\rm 0}^b$ & $T_{\rm 1}^b$ & $r_{\rm tc}^b$ & $r_{\rm 200}$ & $T_{\rm 0.1r_{200}}$ & $n_{\rm e,~0.1r_{200}}$\\
         & (keV) & (keV) & (keV)     & (kpc) & (Mpc)& (keV) & ($10^{-3}$ cm$^{-3}$)\\
\hline
AWM 4 &	$--$	&	$2.34 \pm 0.17$	&	$2.81 \pm 1.37$	&	$21.62 \pm 7.21$ & $0.90 \pm 0.05 $ 	& $2.72 \pm0.05 $	& $2.83 \pm0.32 $\\
ESO 306017	&	$--$	&	$2.14 \pm 0.09 $	&	$3.53\pm 0.87$	&	$143.56 \pm 52.91$ & $1.07 \pm0.03 $ 	& $2.97 \pm0.19 $ & $2.65 \pm 0.12 $\\
NGC 1132 	&	$1.06 \pm 0.01$ &	$--$	&	$--$	&	$--$  & $0.62 \pm0.008 $ 	& $1.04 \pm0.01 $ & $1.40 \pm 0.38 $	\\
NGC 1550	& $--$  & $1.04 \pm 0.02$	&	$1.95\pm 0.21$	&	$35.22 \pm 4.76$ & $0.42 \pm0.02 $ 	& $1.31\pm0.009 $ & $6.78 \pm 0.72$ \\
NGC 6482	&	$--$	& $0.31\pm0.04$	&	$2.29\pm0.28$	&	$4.5\pm0.14$  & $0.53 \pm 0.004 $ 	& $0.65 \pm0.01 $ & $1.08 \pm 0.13$ \\
NGC 741		&	$0.99 \pm 0.02$	&	$--$	&	$--$	&	$--$ & $0.64 \pm0.006$ 	& $1.05 \pm0.02 $ & $1.18 \pm 0.14$ \\
RX J1340.6+4018	&	$1.23 \pm 0.06 $	&	$--$	&	$--$	&	$--$ & $0.64 \pm 0.02 $ 	& $1.15 \pm0.14 $ & $3.71 \pm 0.56$ \\
RX J1416.4+2315	& $4.23 \pm 0.43$ & $--$ & $--$ & $--$ & $1.21 \pm 0.06$ 	& $4.00 \pm0.62 $ & $2.77 \pm 0.14$ \\
SDSS J0150-1005	&	$5.61 \pm 0.52$	& $--$	&	$--$	&	$--$ 	& $1.55 \pm0.11$ 	& $5.28 \pm0.46 $ & $4.63 \pm 0.75$ \\
SDSS J1720+2637	& $--$	& $4.23 \pm 0.32$ & $13.62 \pm 3.65$ & $298.46 \pm 85.94$ & $1.51 \pm 0.08$  & $4.47 \pm 0.49$ & $9.63 \pm0.24 $	\\
\hline\hline
\end{tabular}
\end{center}
\tablecomments{0.9\textwidth}
{
\scriptsize
$^a$ {Global temperatures of five FSs with insufficient counts, which are extracted from a circular region with the center on the peak of the emission corresponding to the maximum measured extent of the X-ray emission.} $^b$ {Best-fit model parameters for deprojected temperature profiles of five FSs with sufficient counts, and we plot the best-fit deprojected temperature profiles as solid lines in fig. \ref{fig_tprofile}.}
}
\end{table}

Due to the limited counts, it's insufficient to extract the deprojected temperature profiles of other five FSs. Therefore, we obtain the global temperature of these five FSs by extracting global ACIS spectra from a circular region with the center on the peak of the emission corresponding to the maximum measured extent of the X-ray emission. The gas are also modeled with an APEC component plus a power-law component, both subjected to a common absorption as above.  Best-fit global temperatures $T_{\rm g}$ of these five members are also listed in Table \ref{table_properties}.

\subsection{Surface Brightness Profile and Gas Density}
\begin{figure}
     \begin{center}
       \includegraphics[width=.8\textwidth]{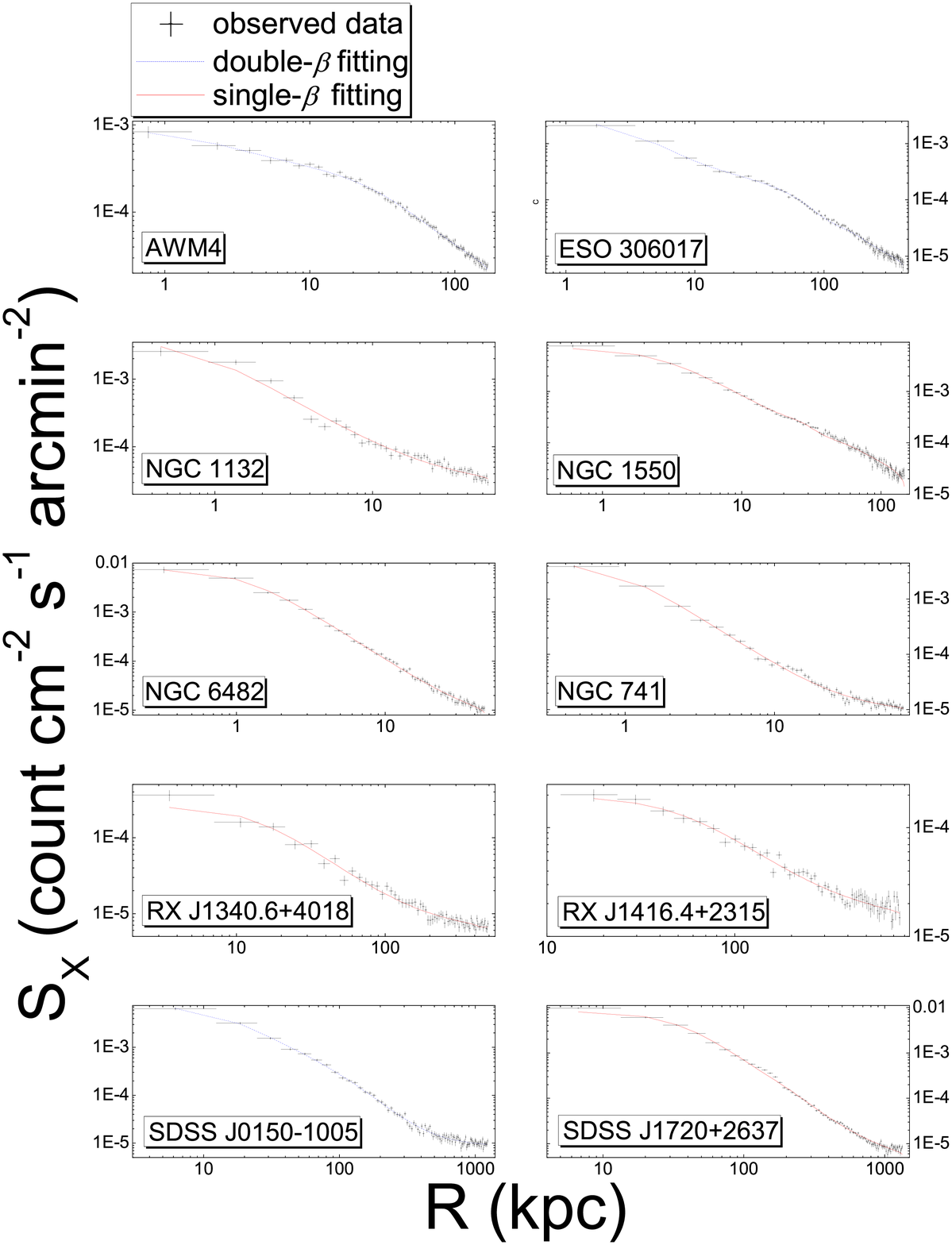}
       \caption{Exposure-corrected SBPs extracted from a series of annuli regions centered on the X-ray emission peak of the FSs. R is the projected distance. The red  solid lines and  blue dash lines are corresponding to the best-fit single-$\beta$ and double-$\beta$ SBPs, respectively. 
       }\label{fig_sbp}
     \end{center}
   \end{figure}

The \chandra\ images of members in the FS sample all exhibit relaxed, regular and symmetry morphology. For every FS, the peak of the X-ray emission consists with the centroid of the cD galaxy within $1^{\prime\prime}$.  We exclude all the visual point sources of FSs, which could be detected at the confidence level of $3\sigma$ by the CIAO tool ${celldetect}$. The X-ray radial surface brightness profiles (SBPs) are extracted from {a series of annuli regions centered on the X-ray emission peak and extended near the boundary of detect CCDs. The energy band of SBPs are restricted to} {the ACIS images in} $0.7-7.0$ keV {band} with exposure corrected. {Such extracted SBPs are shown in Fig. \ref{fig_sbp}.}
\begin{table}[h!!!]
\caption{Best-fit Density Model Parameters}\label{table_ne}
\begin{center}
\tiny
\begin{tabular}{clccccrcc}
\hline\hline
Object name & $ \beta (\beta_{\rm 1})$& $n_{\rm 0} (n_{\rm 1})^a$& $r_{\rm c} (r_{\rm c1})^a$  & $\beta_{\rm 2}$ & $n_{\rm 2}$ & $r_{\rm c2}$& $\chi^2/{\rm dof}^b$ \\
        & &($10^{-2}$ cm$^{-3}$) & (kpc) & &($10^{-2}$ cm$^{-3}$)     & (kpc) &   \\
\hline
AWM 4	&$0.56 \pm 0.01$ &	$4.00 \pm 0.20$	&	$1.34 \pm 0.19$	   &$0.48 \pm 0.01$&	$0.61 \pm 0.04$	&	$31.28  \pm  1.35$ &$113.14/104$	\\
ESO 306017	&	$0.56\pm 0.01$	&	$7.24 \pm 0.18$	&	$3.18 \pm 0.14 $	&	$0.46\pm 0.01$  &	$0.38\pm 0.01$	&	$56.06 \pm 116$ & $117.31/114$ \\
NGC 1132 	&	$0.40 \pm 0.01$	&	$14.82 \pm 2.28$ &	$0.70 \pm 0.12$	&	$--$  &	$--$	&	$--$ & $94.99/56$	\\
NGC 1550	&	$0.35 \pm 0.01$	& $9.35\pm0.83$  & $2.16 \pm 0.03$	&	$--$  	&	$--$   &	$--$ & $119.47/114$	\\
NGC 6482	&	$0.50\pm 0.01$	&	$14.71 \pm 0.90$	& $1.13 \pm 0.06$	&	$--$ &	$--$	&	$--$	& $78.43/68$  \\
NGC 741	&	$0.46 \pm 0.01$	&	$22.32 \pm 1.41$	&	$0.83\pm 0.05$	&	$--$ &	$--$	&	$--$ & $118.66/76$	\\
RX J1340.6+4018	&	$0.42 \pm 0.01$	&	$1.44 \pm 0.14$	&	$13.80 \pm 1.40$	&	$--$ &	$--$	&	$--$ & $68.97/66$	\\
RX J1416.4+2315	&	$0.42 \pm 0.01$	&	$0.50 \pm 0.02$	&	$57.92 \pm 1.40$	&	$--$	&	$--$ &	$--$ & $77.17/59$ \\
SDSS J0150-1005	&	$0.65 \pm 0.02$	&	$5.40 \pm 0.41$	& $15.48 \pm 1.59$	&	$0.64 \pm 0.01$ &	$0.52 \pm 0.01$	&	$107.43  \pm  14.75$ & $97.42/94$	\\
SDSS J1720+2637	&	$0.54 \pm 0.01$	&	$4.88 \pm 0.07$	&	$36.11 \pm 0.54$	&	$--$ &	$--$	&	$--$ & $169.36/96$	\\
\hline\hline
\end{tabular}
\end{center}
\tablecomments{0.9\textwidth}
{
\scriptsize
$^a$ {For the single-$\beta$ fitted FS these two columns represent $n_{\rm 0}$ and $r_{\rm c}$, respectively, and for the double-$\beta$ fitted FS these two columns represent $n_{\rm 1}$ and $r_{\rm c1}$, respectively.} $^b$ {Reduced Chi-square value for the best fit of single-$\beta$ or double-$\beta$ model.}
}
\end{table}

Assuming the spherical symmetry, we deprojected the SBP to 3-dimensional electron number density profile $n_{\rm e}(r)$ of the ICM with standard ``$onion-skin$'' method (Kriss et al. 1983). {During the deprojection, we assume five FSs are isothermal at the best-fit global temperatures $T_{\rm g}$ listed in Table \ref{table_properties}, including NGC 1131, NGC 741, RX J1340.6+4018, RX J1416.4+2315 and SDSS J0150-1005. For the other five FSs, temperature profiles shown in Fig. \ref{fig_tprofile} are applied. We believe that uncertainties from our choice on the gas temperature could not significantly affect resultant electron density profiles, because Kriss et al. (1983) has pointed out that the deduced density profile is actually very insensitive to temperature.} 

\begin{figure}
     \begin{center}
       \includegraphics[width=.8\textwidth]{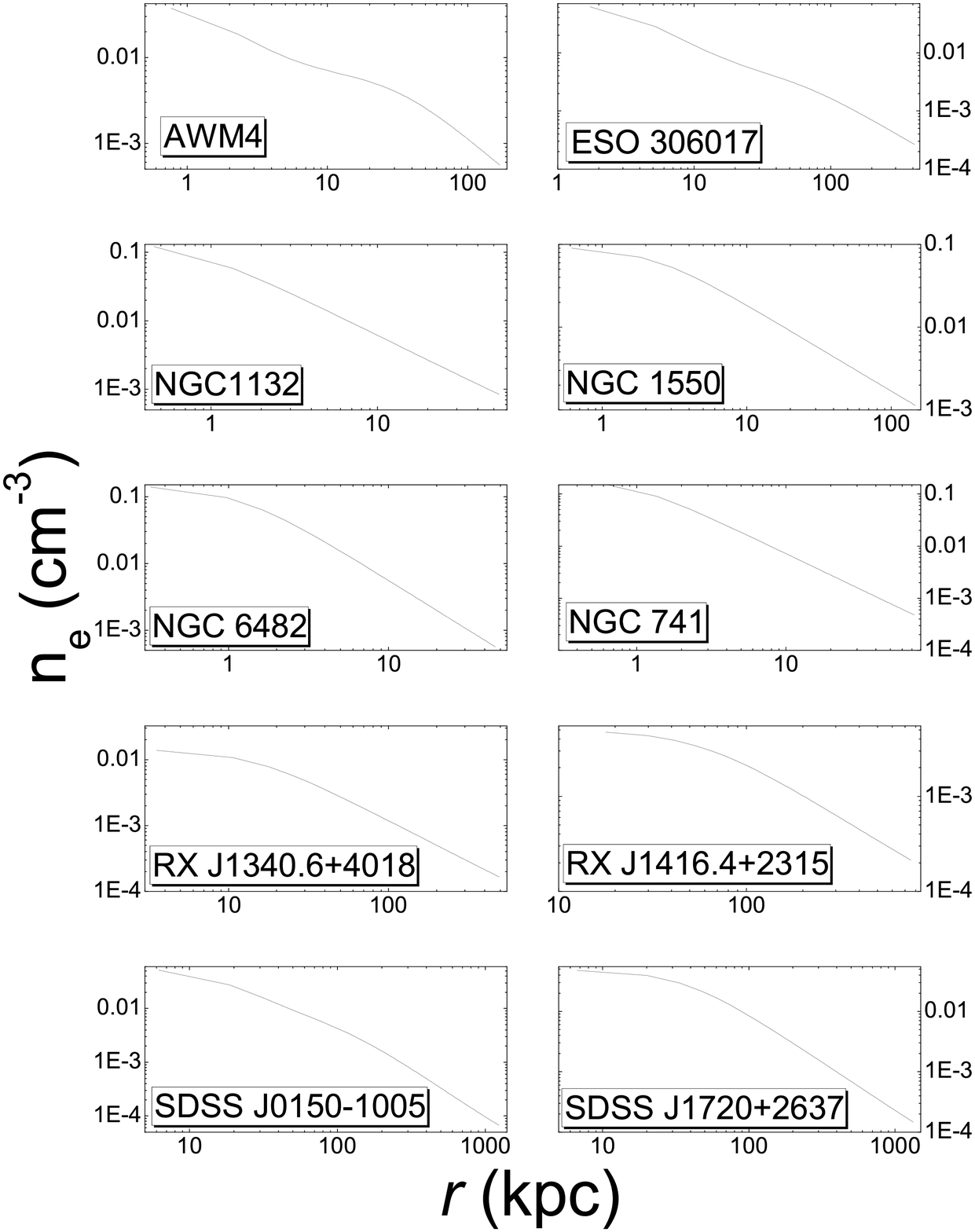}
       \caption{Gas density profiles obtained from the deprojection of the SBPs.
       }\label{fig_ne}
     \end{center}
   \end{figure}

With the single-$\beta$ model, we fit the $n_{\rm e}(r)$ of seven FSs as:
\begin{equation}
n_{\rm e}(r) = n_{\rm 0}[1+(r/r_{\rm c})^2]^{-1.5\beta}+n_{\rm bkg},
\end{equation}
where $r$ is the 3-dimensional radius, $n_{\rm 0}$ corresponds to the normalization, $r_{\rm c}$ is the core radius, $\beta$ is the slope and $n_{\rm bkg}$ is the background. The single-$\beta$ model fails on other three FSs. Because these FSs have an central emission beyond the single-$\beta$ model, which is more appropriate to describe the SBPs of outer regions. To solve this problem, we use double-$\beta$ model to describe the spatial distribution of $n_{\rm e}(r)$ in the other three FSs:
\begin{equation}
n_{\rm e}(r) = n_{\rm 1}[1+(r/r_{\rm c1})^2]^{-1.5\beta_{\rm 1}}+n_{\rm 2}[1+(r/r_{\rm c2})^2]^{-1.5\beta_{\rm 2}}+n_{\rm bkg}.
\end{equation}
We list the best-fit parameters of $n_{\rm e}(r)$ in Table \ref{table_ne}. {The best-fit electron number density profiles and SBPs are presented in Fig. \ref{fig_ne} and Fig.\ref{fig_sbp}, respectively.}

%

\section{Hot Gas Properties}
In this section, we present hot gas properties of FSs, e.g. the relations between gravitation mass, X-ray luminosity ($L_x$), gas fraction ($f_{\rm gas}$), entropy and system temperature ($T$). The spectra temperatures within $0.1r_{\rm 200}$ region are used as the system temperature $T$. First of all, we determine the radius $r_{200}$, $r_{500}$ and $r_{2500}$, within which the average mass density is 200, 500 and 2500 times of the critical density of the universe at corresponding redshift (we will introduce calculation of mass in ${\S}$ 4.1). Table \ref{table_properties} lists all the virial radius $r_{200}$ of our sample in detail. The total mass within $r_{500}$ is denoted as $M_{500}$, which is taken to examine the $M_{\rm 500}-T$ relation of FSs. The bolometric $L_{\rm X}$ of the FS is taken within $r_{200}$. Then, we derive the gas fraction $f_{\rm gas}$, which is the mass ratio of the hot gas component to the total mass. $f_{\rm gas,~2500}$ means the gas fraction at $r_{2500}$. Finally, we calculate entropies within $0.1r_{200}$ of FSs.

\subsection{$M_{\rm 500}-T$ and $L_{\rm X}-T$ Relations}
In a spherically symmetric system with hydrostatic equilibrium, $M_{\rm tot}(<r)$, the total mass within a given radius, $r$, is given by
\begin{equation}\label{eq.m}
M_{\rm tot}(<r) = {{- r^2 k_{\rm B}} \over {G \mu m_{\rm p} n_{\rm e}(r)}} {{d[n_{\rm e}(r) T(r)]} \over {dr}},
\end{equation}
where $G$ is the universal gravitational constant, $k_{\rm B}$ is the Boltzmann constant, $\mu=0.62$ is the mean molecular weight per hydrogen atom, $m_{\rm p}$ is the proton mass, $n_{\rm e}(r)$ is the electron number profile, and $T(r)$ is the temperature profile. In the $M_{\rm tot}(<r)$ calculation, all the gas temperature profiles, $T(r)$, are directly taken from Table \ref{table_properties}, ${\S}$ 3.1. $n_{\rm e}(r)$is obtained from the best-fitting of the X-ray surface brightness profile described in Table \ref{table_ne}. ${\S}$ 3.2. In detail, we use $T(r)$ as the spectral modeling of the five FSs with sufficient counts, and assume a constant $T(r)$ for the rest five FSs, where one can only get the global temperatures $T_{\rm g}$. With Eq. (\ref{eq.m}) and 1000 Monte-Carlo simulations, $M_{\rm 500}$ and its errors are readily derived.

The luminosities, $L_{\rm X}$ of the FSs within $r_{200}$ is given by $L_{\rm X}=\iiint \Lambda n_{\rm e} n_{\rm H} dV$, where $\Lambda$ is cooling function and $n_{\rm H}$ is the proton number density. We calculate the integral of electron number density profile to get the bolometric $L_{\rm X}$. The error in the $L_{\rm X}$ is obtained from the Poisson error in the X-ray count rate.

\begin{figure}
     \begin{center}
       \includegraphics[width=.8\textwidth]{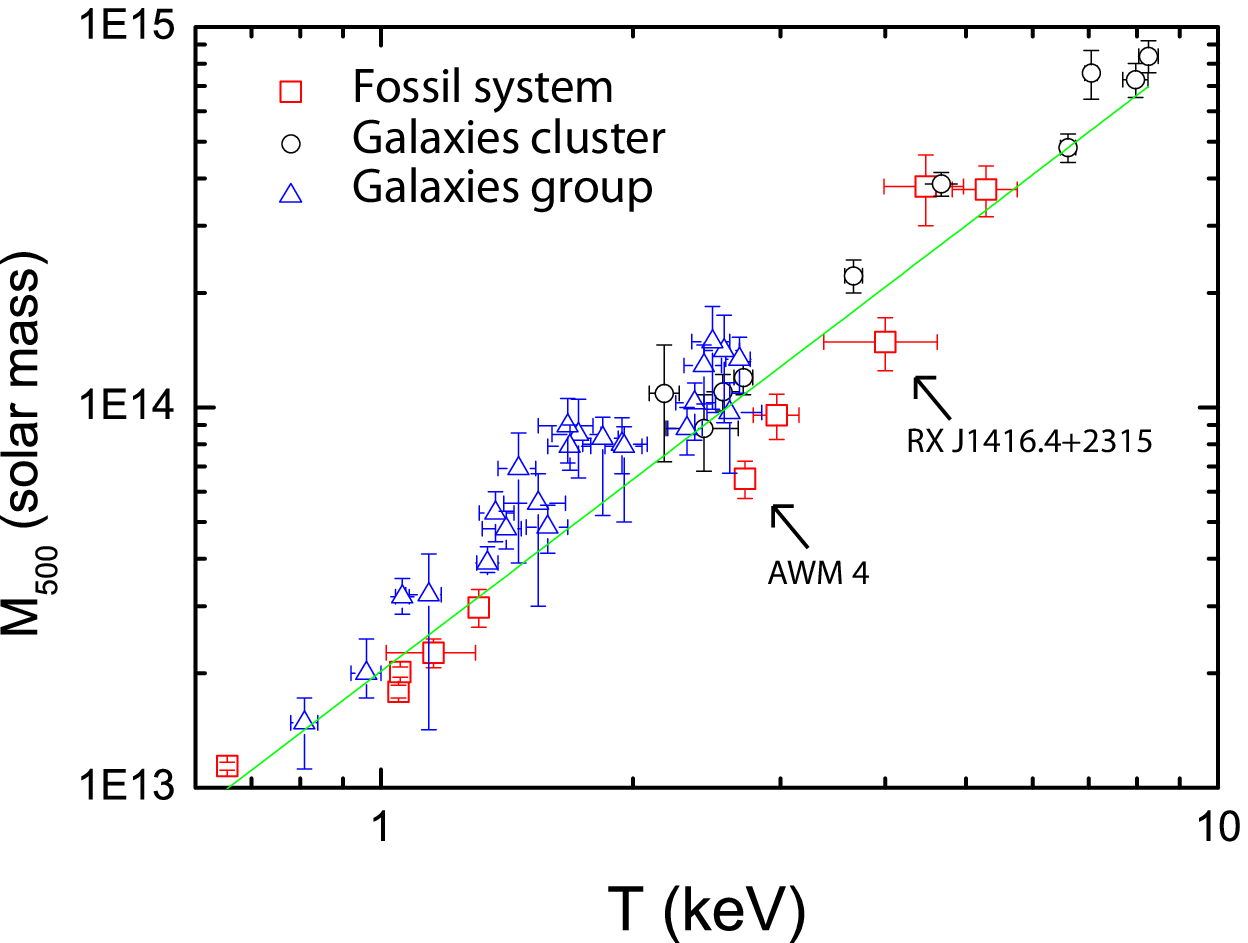}
       \caption{$M_{\rm 500}-T$ relation for FSs compared with ordinary galaxy groups and clusters. The red squares represent FSs, black circles correspond to galaxy clusters derived from Arnaud et al. (2005), and blue triangles correspond to galaxy groups derived from Sun et al. (2009).  {Two fossil clusters in the sample, AWM 4 and RX J1416.5+2315, are marked}. The green solid line is the best linear fit to all the date with slope $1.67 \pm 0.04$.
       }\label{fig_mt}
     \end{center}
   \end{figure}

\begin{figure}
    \begin{center}
     \includegraphics[width=.8\textwidth]{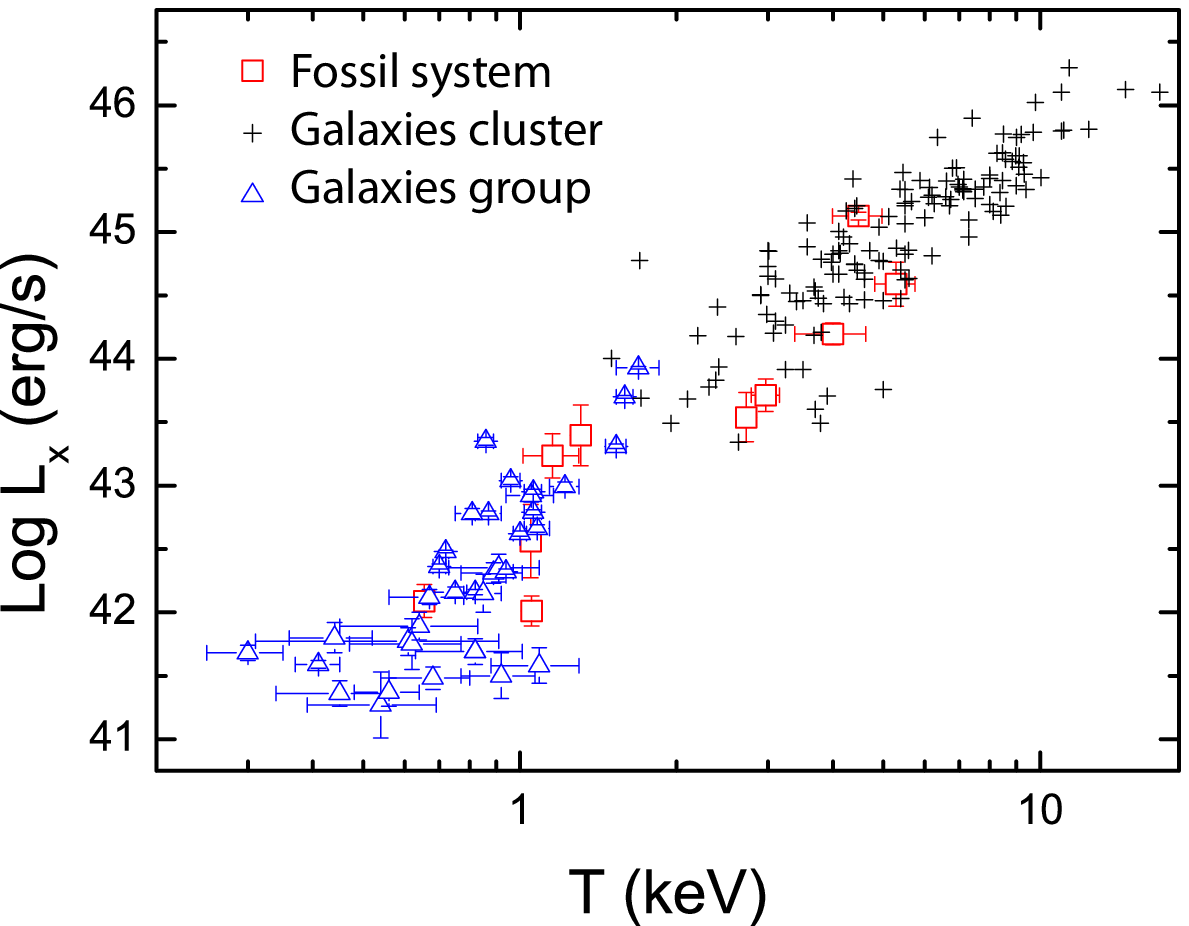}
     \caption{$L_{\rm X}-T$ relation of FSs compared with the non-fossil systems. Red squares represent FSs, black crosses correspond to galaxy clusters derived from Wu et al. (1999), and blue triangles correspond to galaxy groups derived from Helsdon et al. (2001). The FSs sample falls on the $L_{\rm X}-T$ relation of the non-fossil systems.
       }\label{fig_lt}
      \end{center}
      \end{figure}

Fig. \ref{fig_mt} shows the $M_{\rm 500}-T$  relation of FSs compared with galaxy groups and clusters from Sun et al. (2009) and Arnaud et al. (2005), respectively. The slope of the best linear fit to all the data is $1.67 \pm 0.04$, which consists with the value of previous work $1.78 \pm 0.09$ within the errors (Finoguenov et al. 2001). Fig. \ref{fig_lt} presents the $L_{\rm X}-T$ relation of FSs compared with non-fossil galaxy groups and clusters from the sample of Helsdon et al. (2000) and  Wu et al. (1999), respectively. The $M_{\rm 500}-T$ and $L_{\rm X}-T$ relations of FSs exhibit no obvious deviate from  non-fossil galaxy groups and clusters.

\subsection{Gas Fraction of FSs}
\begin{figure}
    \begin{center}
     \includegraphics[width=.8\textwidth]{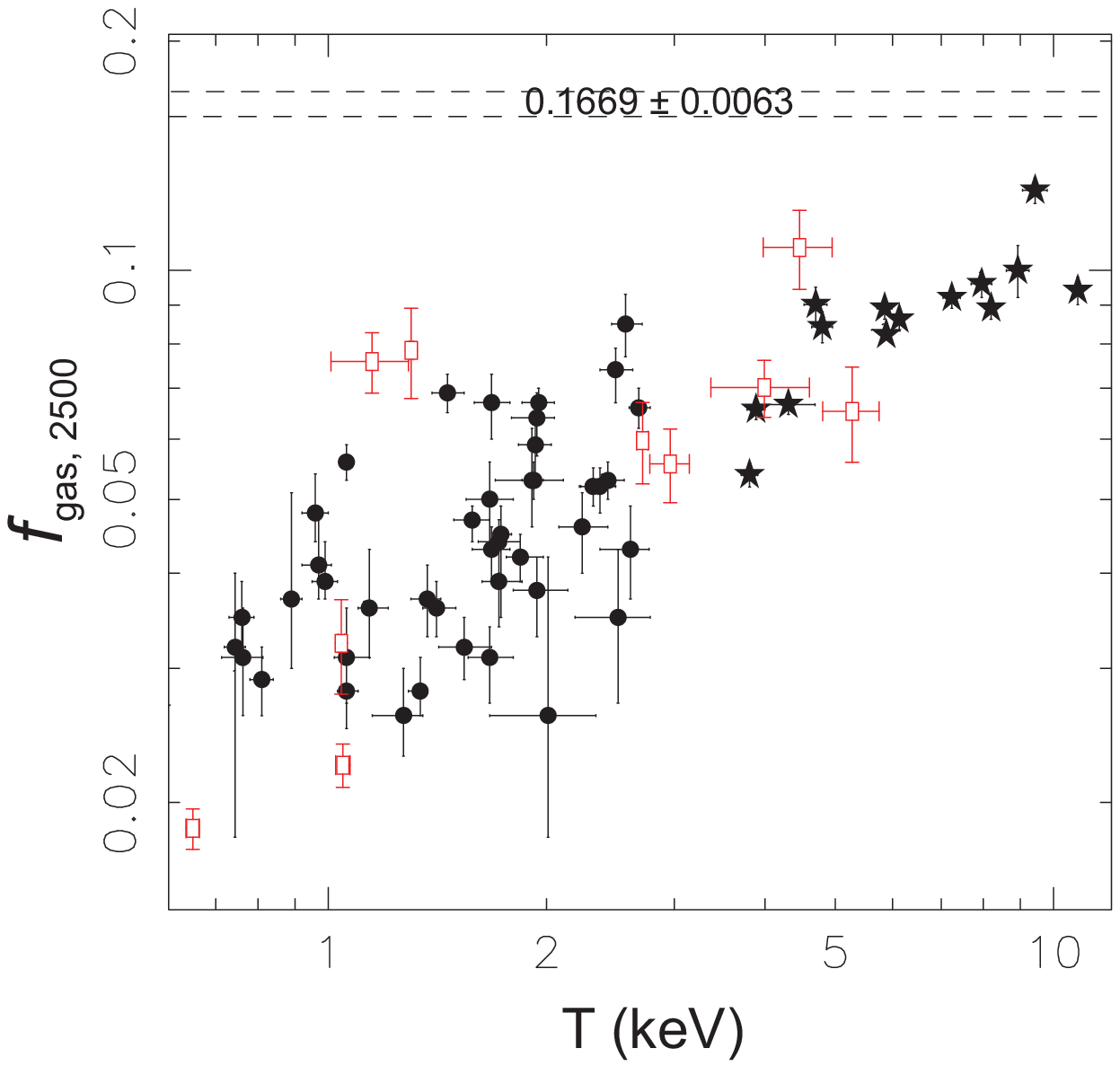}
     \caption{$f_{\rm gas,~2500}$ of the FSs sample compared with ordinary systems. Red squares represent the FSs, full circles correspond to 43 galaxy groups from Sun et al. (2009), and full stars correspond to 14 galaxy cluster from Vikhlinin et al. (2006; 2009). Two dashed lines enclose the 1 $\sigma$ region of the universal baryon fraction derived from the $WMAP$ 5-year data combined with the data of the TyPE Ia supernovae and Baryon Acoustic Oscillations (0.1669 $\pm$ 0.0063, Komatsu et al. 2009).
      }\label{fig_frac}
      \end{center}
      \end{figure}

 With $n_{\rm e}(r)$ from $\S$ 3.2, we calculate the $f_{\rm gas}$ for FSs at $r_{200}$, $r_{500}$ and $r_{\rm 2500}$. The gas fractions increase with radius in FSs, which is consistent with the ordinary systems (Gastaldello et al. 2007). Typically, we present the $f_{\rm gas,~2500}-T$ relation in Fig. \ref{fig_frac} compared with 43 galaxy groups (Sun et al. 2009) and 14 galaxy clusters (Vikhlinin et al. 2006; 2009). In our sample, seven higher temperature FSs show a nearly constant gas fraction with mean  value of  $f_{\rm gas,~2500}$, $0.073 \pm 0.007$, while other three lower temperature FSs exhibit the decreased $f_{\rm gas,~2500}$. The $f_{\rm gas,~2500}-T$ relation in FSs is consistent with previous works, which have reported that $f_{\rm gas,~2500}$ remains quite constant over higher temperature galaxy systems, yet detected a decrease in the low mass, low temperature end (Gastabdello et al. 2007; Dai et al. 2010). 


\subsection{Entropy Within $0.1r_{\rm 200}$}
\begin{figure}
    \begin{center}
     \includegraphics[width=.8\textwidth]{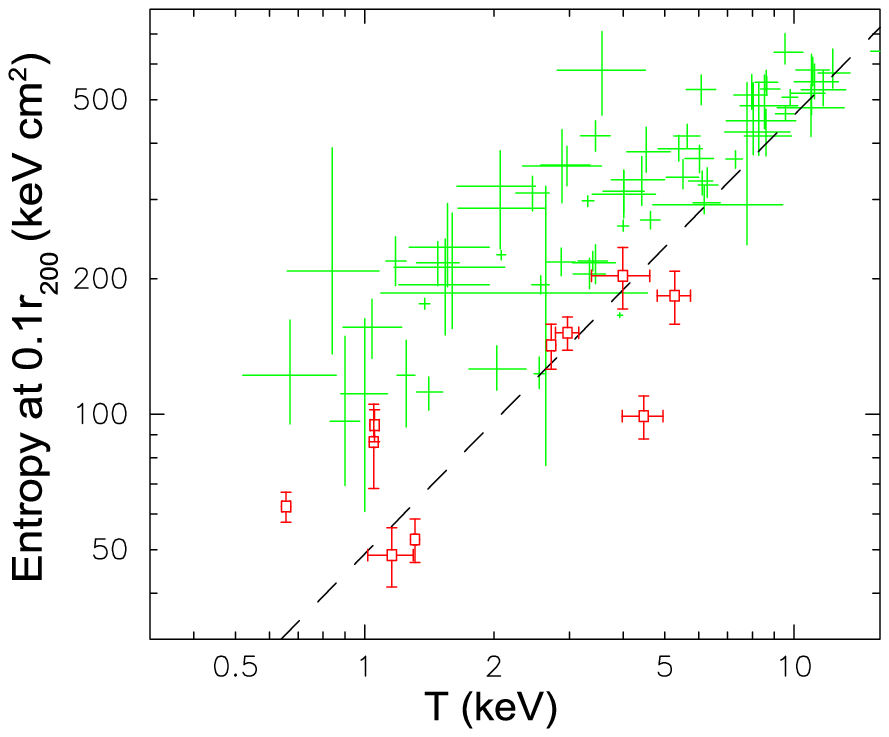}
     \caption{ Gas entropy at $0.1r_{\rm 200}$ as a function of system temperature for FSs (red boxes) compared with non-fossil galaxy groups and clusters (green crosses) from Ponman et al. (2003). The dashed line is the extrapolation of self-similarity normalized to the mean entropy of the hottest 8 clusters in Ponman et al. (2003).
      }\label{fig_st}
      \end{center}
      \end{figure}

The gas entropy is defined as
\begin{equation}
S= T/n_{\rm e}^{2/3},
\end{equation}
where $n_{\rm e}$ is the electron number density. We calculate the gas entropy at $0.1r_{\rm 200}$, $S_{\rm 0.1r_{200}}$. Because $0.1r_{200}$ is very close to the center, we can avoid the shock-generated entropy and thus enhance the sensitivity to any additional entropy. Fig. \ref{fig_st} shows the entropies of FSs compared with Ponman et al. (2003). The entropies of FSs are systematically lower than those of the non-fossil galaxy groups, and the $S_{0.1r_{200}}-T$ plots of FSs are roughly around the dashed line in Fig. \ref{fig_st}, which has self-similar slope of 1, normalized to the mean entropy of the hottest 8 clusters in Ponman et al. (2003). We also list $n_{\rm e}$ within $0.1r_{\rm 200}$ regions of FSs in Table \ref{table_properties}, which are $\sim 10^{-3}$ cm$^{-3}$ at same order as galaxy clusters, but higher than those of galaxy groups.

\section{DARK MATTER HALO STRUCTURES}

Suggested by the high-resolution N-body simulations, Navarro, Frenk, \& White (1996; hereafter NFW) model is used to describe the mass profile of all the dark halos:
\begin{equation}
\rho(r) = {\rho_{\rm 0} ta_{\rm c} \over (r/r_{\rm s})(1+r/r_{\rm s})^2},
\end{equation}
where $\rho$ is the mass density, $r_{\rm s}$ is the scale radius, $\rho_{\rm 0}$ is the critical density of the universe, and $ta_{\rm c}$ is the characteristic density. With the NFW model, we can obtain the following integrated profile of a spherical mass distribution,
\begin{equation}\label{eq_m}
M_{\rm tot}(<r) = 4 \pi ta_{\rm c} \rho_{\rm 0} r_{\rm s}^3[\ln(1+{r \over r_{\rm s}}) - {r \over r+r_s}].
\end{equation}
We fit the NFW mass profile, eq. (\ref{eq_m}), to our  X-ray derived total mass profile of FSs in $(0.1-1)r_{200}$ to determine $r_{\rm s}$ and $ta_{\rm c}$.

\begin{figure}
    \begin{center}
     \includegraphics[width=.8\textwidth]{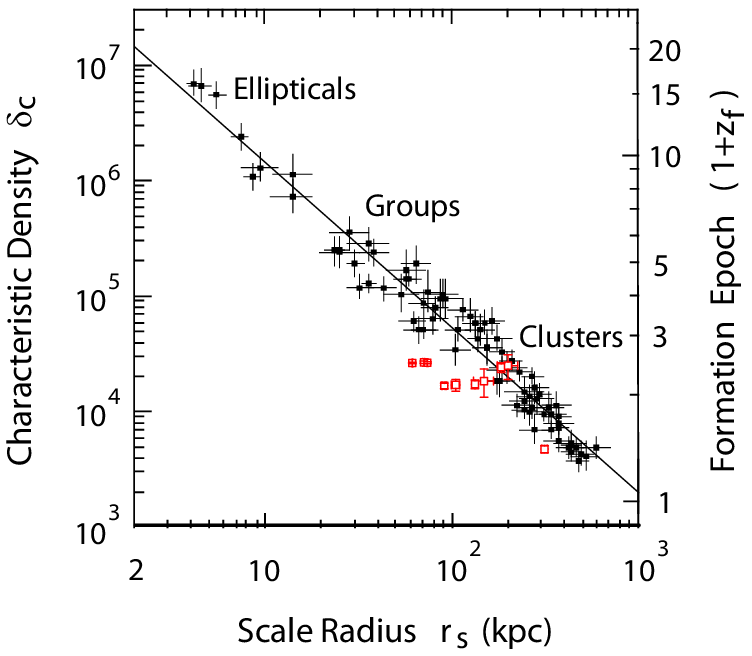}
     \caption{The $ta_{\rm c}-r_{\rm s}$ relation. Red squares show FSs, black filled squares represent ellipticals, groups and clusters from Sato et al. (2000). Dark matter halos of FSs assembled approximately at the epoch of $1 + z_{\rm f} \simeq 2-3$
      }\label{fig_dcrs}
      \end{center}
      \end{figure}

Fig. \ref{fig_dcrs} shows the relation between $r_{\rm s}$ and $ta_{\rm c}$ of FSs. The FSs show a similar correlation between $ta_{\rm c}$ and $r_{\rm S}$ as in Sato et al. (2000), which indicates the self-similarity of dark matter halos. According to the range of $r_{\rm s}$ and $ta_{\rm c}$, the sizes of dark matter halos of FSs are more similar to those of galaxy clusters.


Since the density profile of dark halo is rearranged only by a major merger, it indicates when the last merger took place: each $ta_{\rm c}$ in NFW model corresponds to a forming epoch of dark halo. Therefore, dark matter halos of FSs were assembled approximately at $1 + z_{\rm f} \simeq 2-3$ as shown in Fig. \ref{fig_dcrs}. 
\section{DISCUSSION}

In $\S$ 4 and $\S$ 5, we present hot gas properties and dark matter halo structures of FSs. The gas properties of FSs are not physically distinct from ordinary systems, but their special birth places make FSs as a distinct galaxy system type. In detail, we discuss this result as follows.  

First of all, according to Figs. \ref{fig_mt}-\ref{fig_lt}, the FSs have the similar $M_{\rm 500}-T$ and $L_{\rm x}-T$ relations to normal groups and clusters, which means the gas properties of FSs are not a physically distinct from normal objects. In the standard picture of hierarchical structure formation, groups comprising a handful of galaxies merge through gravity to form large clusters of hundreds of galaxies. The $M_{\rm 500}-T$ and $L_{\rm X}-T$ relations also indicate that FSs follow the standard picture, and belongs to a common class of virialized systems in the hierarchical structure forming universe. We also present the $f_{\rm gas,~2500}-T$ relation of FSs in Fig. \ref{fig_frac}, which agrees with current observations(Gastabdello et al. 2007; Dai et al. 2010). In simulation, gas fraction directly related to the strength of cooling, star formation and AGN in the galaxy group and cluster (e.g., Kravtsov et al. 2005; Puchwein et al. 2008). Therefore, our result indicates these non-gravitational events in FSs may be as active as ordinary systems. This is supported by the study on the level of AGN activity in fossil systems (Hess et al. 2012). In sum, the gas properties of FSs are not physically distinct from ordinary systems.


Secondly, we examine the entropies within $0.1r_{200}$ of our sample. As show in Fig. \ref{fig_st}, the entropies of FSs are systematically lower than ordinary galaxy groups or clusters. If FSs evolve from the normal galaxy groups through a fast and efficient process of merging, the merger shocks during this process would generate entropy, so that the FSs have higher entropy than normal galaxy groups (Tozzi et al. 2000). This is opposite to our result. Therefore, FSs are unlikely formed through a fast and  efficient merging of normal galaxy groups.

On the other hand, the $S_{0.1r_{200}}-T$ relation follows the self-similar extrapolation of rich-cluster entropies. In previous work, entropy excess in the inner regions of galaxy groups and clusters are explained by the pre-heating scenario, which suggests the existence of a universal entropy floor. The entropy excess is greater in lower temperature systems (Ponman et al. 1999; Finoguenov et al. 2002). As presented in Fig. \ref{fig_st}, the entropy excess in FSs are less than ordinary galaxy groups, but at the same level of ordinary galaxy clusters. The gravitational collapse of galaxy clusters happened earlier than the galaxy groups, which makes the entropy excess is greater in cooler systems: in hotter clusters the gravitational collapse of the system generates entropies in excess of the floor value established by preheating, but in cooler systems it is preserved during collapse, and prevents the gas from collapsing to high central densities. (Ponman et al. 1999). Therefore, FSs may collapse as early as galaxy clusters, and FSs are more similar to galaxy clusters in the preheating history.

Finally, we study dark matter halo structures of FSs by fitting its mass profile with the NFW model. As shown in Fig. \ref{fig_dcrs}, FSs have similar dark matter halo structures to the ordinary objects (Sato et al. 2000). The range of FS $r_{\rm s}$ and $ta_{\rm c}$ is more close to that of galaxy clusters, which indicates that FSs and galaxy clusters may have the nearly same size dark matter halos. According to Sato et al. (2000), dark matter halos of FSs were assembled approximately at $1 + z_{\rm f} \simeq 2-3$, which means the forming epoch of FSs is later than the galaxy groups, but the same time as galaxy clusters. This is consistent with the simulation of  D{\'{\i}}az-Gim{\'e}nez et al. (2008), which points out that the first-ranked galaxies in fossil systems merged later than their non-fossil-system counterparts.

FSs share many common characteristics with galaxy clusters. Their gravitational collapse times and dark matter halo sizes are nearly the same, and also the $n_{\rm e}$ within $0.1r_{\rm 200}$ regions of FSs are $\sim 10^{-3}$ cm$^{-3}$ at same order as galaxy clusters. Why the FSs have large magnitude gaps which distinct from ordinary galaxy cluster? We prefer the answer that the birth places of FSs may be isolated and deficient in $L^*$ galaxies, so that the FSs exhibit large magnitude gaps compared with galaxy clusters. All in all, we would conclude that the special birth place of the FS makes it as a distinct galaxy system type.

\section{SUMMARY}
By using \chandra\ X-ray observations of a sample of ten FSs, we study their $M_{500}-T$, $L_{\rm X}-T$ and $f_{\rm gas,~2500}-T$ relations, and find the hot gas properties of FSs are not physically distinct from ordinary systems. By analyzing the gas densities, entropies and dark matter halos of FSs, we also find FSs share common characteristics with galaxy clusters, such as they have almost the same gas densities, gravitational collapse times and dark matter halo sizes. Therefore, we prefer the scenario that birth place of FS may be isolated and deficient in $L^*$ galaxies, which makes it as a distinct galaxy system type. 



\label{lastpage}
\end{document}